\documentclass{article}

\usepackage{fullpage}

\usepackage{amsfonts}       
\usepackage{amssymb, amsmath, amsthm}
\usepackage{graphicx}
\usepackage[caption=false]{subfig}
\usepackage{setspace}

\theoremstyle{plain}

\theoremstyle{definition}

\theoremstyle{remark}

\title{Subgrid-scale parametrization of unresolved scales in forced Burgers equation using Generative Adversarial Networks (GAN)}

\author{
  Jeric Alcala\thanks{
  Dept. of Mathematics,
  University of Houston,
  Houston, TX 77204,
  \texttt{jsalcala@math.uh.edu}}
   \and
 Ilya Timofeyev\thanks{
  Dept. of Mathematics,
  University of Houston,
  Houston, TX 77204,
  \texttt{ilya@math.uh.edu}}
 }
 
\begin{document}
\maketitle

\begin{abstract}
Stochastic subgrid-scale parametrizations aim to incorporate effects of unresolved processes in an effective model by sampling from a distribution usually described in terms of resolved modes. This is an active research area in climate, weather and ocean science where processes evolved in a wide range of spatial and temporal scales. In this study, we evaluate the performance of conditional generative adversarial network (GAN) in parametrizing subgrid-scale effects in a finite-difference discretization of stochastically forced Burgers equation. We define resolved modes as local spatial averages and deviations from these averages are the unresolved degrees of freedom. We train a Wasserstein GAN (WGAN) conditioned on the resolved variables to learn the distribution of subgrid flux tendencies for resolved modes and, thus, represent the effect of unresolved scales. The resulting WGAN is then used in an effective model to reproduce the statistical features of resolved modes. We demonstrate that various stationary statistical quantities such as spectrum, moments, autocorrelation, etc. are well approximated by this effective model.
\end{abstract}



\section{Introduction}


Many physical processes in mathematics, engineering and other applied sciences are represented by differential equations. In practical numerical applications partial differential equations are discretized in space and time. Often finite-difference or finite-volume numerical schemes are used for
spatial discretization which results in a system of ordinary differential equations or difference equations. However, in doing so, model error are introduced and some of the physical processes may not be fully resolved on the defined spatial grid. Either, some processes evolve on a much smaller spatial (subgrid) scale compared to the chosen mesh size or there are inherent properties of the system (e.g. nonlinearity, chaotic behavior) that cannot be captured by the chosen discretization. The problem of incorporating the approximate effects of unresolved processes or sub-grid scale processes in an effective model is called parametrization (or sub-grid scale parametrization, in our context).

Developing accurate and fast parametrizations has been an active research area in ocean, climate and atmospheric science for many decades. For deterministic parametrization schemes a function relating unresolved modes with every state of the resolved modes is constructed either by using known physics or existing high-resolution data. This leads to incorporating the {\it mean} impact of subgrid processes on resolved modes. To be realistic, some measure of variability of the impacts of these processes must be taken into account as proposed by Hasselman \cite{hassel} in a visionary paper in 1976. Introduction of randomness leads to stochastic parametrizations.                    

In recent decades, there have been a considerable number of studies focusing on stochastic parametrizations. Many methods have been developed which aim to parametrize effects of sub-grid scales by incorporating stochastic terms with the goal of reproducing the statistical behavior of resolved scales. To mention a few, there are techniques which use regression models (e.g. \cite{wilks05, kkg05}), add stochastic perturbations to the mean parametrization terms \cite{sppt1, sppt2}, utilize data-driven methods such as NARMAX fitting \cite{chorin15, narmax2}, or use conditional Markov chains \cite{markov1, markov2, markov3}.
There are also more rigorous approaches (see recent review paper \cite{resseguier20}) including 
stochastic mode reduction strategy exploiting scale separation between 
the resolved and sub-grid scale processes \cite{mtv01, mtv03, seam1, low1, low2}, 
reduction techniques based on the 
Zwanzig-Mori formalism \cite{chh09,Chorin2968,Chorin2002239,hevd10}, 
turbulence closure techniques \cite{frederiksen97,Frederiksen2013,Verkley2016}.
In particular, 
stochastic mode-reduction has been applied recently to reproduce the statistical behavior of local 
spatial averages in the truncated Burgers-Hopf model \cite{dta12, dat12} 
and the Shallow-Water equations \cite{zdat18}.
A more relaxed scheme that relies on weak coupling and uses ideas from response theory has been developed in \cite{resp1}. Another approach uses maximum entropy principle for estimating probability density of unresolved scales \cite{maxent1, maxent2}. 
Stochastic parametrizations have been proven to be extremely useful in weather and climate science for analyzing prediction uncertainty, representing sub-grid processes, capturing response to forcing and inducing regime transitions (see e.g. \cite{bams}).

Machine learning has become increasingly popular in recent years offering possible data-driven solutions for problems in many areas of physical and natural sciences. Machine learning also received increased attention in applied and computational fluid dynamics as well.
One advantage of most machine learning methods is that they do not necessarily require assumptions about the model equations and/or structure of the data. Recently, neural networks (NNs) became one of the most popular area of machine learning and there are several applications of NNs in various models relevant to 
atmospheric research. In particular, NNs were used to represent clouds \cite{rasp} and convection \cite{gordw} in a global circulation model. NN was also used to develop a parametrization for
residual heating and moistening in a global circulation model \cite{bb}.
Various deep learning methods such as reservoir computing and LSTM neural networks were used to predict 
trajectories of chaotic dynamics (e.g. \cite{nn1,nn2}).
Methods used in these papers are classified as supervised learning and results in deterministic parametrizations. 
Here we use generative models to develop NN-based stochastic parametrizations.

Application of generative models is one of the relatively new areas in machine learning. 
These are primarily techniques from unsupervised learning in which a network is trained to produce samples from a given distribution. One popular generative model is {\it generative adversarial networks} (GAN) introduced in \cite{gan}. This model consists of two networks - {\it generator} and 
{\it discriminator}. Given a training set, these two networks compete with each other - 
the generator attempts to produce new samples consistent with the training set, and the discriminator
evaluates them with the goal of distinguishing whether these samples came from the original distribution.
Thus, GAN aims to learn the underlying distribution for the training data in order to produce new samples that a critic (i.e. discriminator) qualifies to be coming from that distribution. 
It is easy to see that using GAN would result in a stochastic parametrization. 
A modified version of this approach was applied to Lorenz 96 model
\cite{ganlorenz}. Recently, an alternative training metric for generative adversarial networks
has been proposed. It particular, it was demonstrated that Wasserstein GAN (WGAN)
might have certain advantages over the traditional GAN during training
\cite{wgan, wgangp}.

In this paper a conditional WGAN is trained to learn the distribution of subgrid processes and then used to generate possible values of unresolved modes based on the current state of the resolved variables. We apply this approach for parametrizing local residuals in a flux discretization of stochastically forced Burgers' equation. Specifically, a WGAN conditioned on resolved modes is trained to produce samples from the distribution of subgrid fluxes to be used in an effective model for local averages. 




The rest of the paper is organized as follows. In section \ref{subflux} we discuss general methodology for 
defining coarse variables and representing the effective equations using flux discretization. In section \ref{sec:burgers}
we discuss application of this methodology to the forced Burgers-Hopf model. In section \ref{sec:wgan} we 
discuss configuration and training of the neural network used to represent the subgrid fluxes.
In section \ref{sec:results} we present numerical results comparing 
stationary statistics of the full and reduced models and discuss performance of the WGAN parametrization with respect to changes in forcing and resolution. Final remarks are presented in section \ref{sec:conc}.

\section{Methods}

\subsection{Subgrid flux parametrization}
\label{subflux}

Assume that a system is described by a high dimensional state vector $\mathbf{u}$ which can be decomposed as $\mathbf{u} = \left( \mathbf{U}, \mathbf{y} \right)$ where $\mathbf{U}$ are the resolved variables and $\mathbf{y}$ are unresolved (subgrid) degrees of freedom. In the context of this paper, vector 
$\mathbf{u}$ of time-dependent variables is obtained by a suitable finite-difference or finite-volume discretization of a partial differential equation. Typically, vector $\mathbf{u}$ is high-dimensional, while 
$\text{dim}(\mathbf{U}) \ll \text{dim}(\mathbf{u})$.

The complete dynamics of $\mathbf{u}$ can be recast as a coupled system for resolved and
unresolved variables
\begin{align}
    \dfrac{d}{dt} \mathbf{U} & = \mathbf{F}(\mathbf{U}) + \mathbf{G}(\mathbf{U,y}), \label{fullparam1} \\
    \dfrac{d}{dt} \mathbf{y} & = \mathbf{f}(\mathbf{y}) + \mathbf{g}(\mathbf{U,y}), \label{fullparam2}
\end{align}
where $\mathbf{F}(\mathbf{U})$ represents the interactions among the resolved modes, 
$\mathbf{G}(\mathbf{U,y})$ represents interactions between the resolved and unresolved variables, etc.
In the context of subgrid scale parametrization developed in this paper, the first equation \eqref{fullparam1} describes the time evolution of large-scale variables on a coarser grid. The term $\mathbf{G}(\mathbf{U},\mathbf{y})$ represents subgrid scale tendencies arising from the interaction among the resolved and unresolved degrees of freedom. In the most general sense, the aim of parametrization is to approximate $\mathbf{G}(\mathbf{U}, \mathbf{y})$ in terms of the resolved modes only so that the resulting closed system approximates, in some sense, the behavior of resolved variables $\mathbf{U}$ in the full dynamics.
This paper is focused on stochastic parametrizations where approximations to $\mathbf{G}$ are drawn from the distribution of subgrid-scale tendencies. In addition, it is necessary to condition the 
distribution of $\mathbf{G}$ on the current values of the large-scale variables,  $\mathbf{U}$.

The method developed in this paper applies to system based on flux difference discretization of partial differential equations that can be written in conservative form. More precisely, consider a state field $u(x,t)$ which can be discretized on a uniform fine mesh $\mathcal{M}_f = \{0,1,2,..., N-1\}$ resulting in a system of ODEs
\begin{equation}
\dfrac{d}{dt}u_i = -\dfrac{F_{i+1/2} - F_{i-1/2}}{\Delta x} + \rho \label{fullu} , 
\end{equation}
where $\Delta x = x_{i+1/2} - x_{i-1/2}$. The value $u_i$ is the numerical approximate for either the value of $u(x,t)$ at the center of interval $[x_{i-1/2}, x_{i+1/2}]$ or for the average $\frac{1}{\Delta x}\int_{x_{i-1/2}}^{x_{i+1/2}} u(x,t) dx$ depending on whether the numerical method used is a finite difference or a finite volume scheme, respectively. 
More applicable to finite volume methods, $F_{i+1/2}$ and $F_{i-1/2}$ are discrete approximations for the flux at the right and left ends of the interval $i$, respectively. 
Here, $\rho$ represents forcing. Equation above represents the full dynamics. Typically, direct numerical simulations (DNS) of the equation \eqref{fullu} 
(same as \eqref{fullparam1}, \eqref{fullparam2})
on a small enough mesh are computationally expensive and it is desirable to obtain a closed-form 
system for variables $\mathbf{U}$ alone.

We define the resolved modes as the mean of $u_i$ over $n$ neighboring high resolution grid points, that is, 
\begin{align}
U_I = \dfrac{1}{n}\sum_{i=nI}^{n(I+1)-1}u_i \label{aveop} 
\end{align}
for each cell $I$ in a coarse mesh $\mathcal{M}_c = \{0,1,2,N/n - 1\}$. The unresolved degrees of freedom are then the residuals or deviations $y_i = u_i - U_{I(i)}$, where index $i$ for $y_i$ and $u_i$ refers to the coarse cell $I$ where fine cell $i$ is located. The idea of using averaging operator is similar to large-eddy simulation (LES) modelling in turbulence. Evidently, some degree of scale separation can be expected between $U_I$ and $y_i$ by this construction, but it might depend on the number of averaged points, $n$. Our goal is to develop a closed effective equation for variables $U_I$ on a coarser mesh of size $n\Delta x$. To this end, we consider coupled dynamical system for $\mathbf{U} = \{U_I, \, I\in \mathcal{M}_c\}$ and $\mathbf{y} = \{y_i, \, i \in \mathcal{M}_f\}$. 
There are various ways to decompose the dynamics in \eqref{fullu} and define fluxes
$\mathbf{F}(\mathbf{U})$ and $\mathbf{G}(\mathbf{U,y})$. However, a
natural way is to use the same flux discretization for 
$\mathbf{F}(\mathbf{U})$ as for the full dynamics, but now applied to the coarse mesh. 
This results in the system
\begin{align}
\dfrac{d}{dt}U_I &= -\dfrac{F_{n(I+1) - 1/2}-F_{nI-1/2}}{n\Delta x} + \rho^{U_I}, \label{fbeslow}\\
\dfrac{d}{dt}y_i & = -\dfrac{F_{i+1/2}-F_{i-1/2}}{\Delta x} + \dfrac{F_{n(I+1) - 1/2}-F_{nI-1/2}}{n\Delta x} + \rho^{y_i}, \label{fbefast} 
\end{align}
where $\rho^{U_I}$ and $\rho^{y_i}$ are the projections of the forcing on the respective modes. We will refer to the coupled system \eqref{fbeslow}, \eqref{fbefast} as the {\it full model}. 
Note that $F_{n(I+1)-1/2}$ is the flux at the right boundary of the coarse cell $I$ and it is a function of both the resolved and unresolved modes. We can express this flux as
\begin{align}
F_{n(I+1)-1/2}(\mathbf{U}, \mathbf{y}) = F_{I+1/2}(\mathbf{U}) + G_{I+1/2}(\mathbf{U}, \mathbf{y}) \label{sgf}
\end{align}
where $G_{I+1/2}$ is referred to as the {\it subgrid flux} at the right boundary of cell $I$. 
Term $F_{I+1/2}(\mathbf{U})$ describes self-interactions of local averages on the coarse mesh using the same discrete flux used in \eqref{fullu}. Physical processes which cannot be adequately resolved with the mesh size $n\Delta x$ are represented by $G_{I+1/2}$.
Also, note that although we write the entire vector $\mathbf{U}$ and $\mathbf{y}$ as predictors, only few neighboring\footnote{Number depends on the stencil used to represent the numerical flux} $U_I$ and $y_i$ are actually used in the right-hand size of the equation for $U_I$. 

We would like to point out that decomposition \eqref{fbeslow}, \eqref{fbefast}, and \eqref{sgf} is a particular case of more
general form of the resolved - subgrid variables decomposition presented in \eqref{fullparam1}, \eqref{fullparam2}. Decomposition in \eqref{fbeslow}, \eqref{fbefast}, and \eqref{sgf} imposes a particular
structure on the right-hand side of the full model, and thus, on the right-hand side of the effective model
for the resolved variables. In particular, the model in 
\eqref{fbeslow}, \eqref{fbefast}, \eqref{sgf} involves subgrid fluxes in an explicit form and, thus, 
the goal is to develop parametrization for the subgrid fluxes
$G_{I+1/2}(\mathbf{U}, \mathbf{y})$. 
Formulation in \eqref{fbeslow}, \eqref{fbefast}, \eqref{sgf}
has several potential advantages. In particular,
(i) subgrid fluxes defined in \eqref{sgf} may contain features that can be lost when net 
flux tendencies in \eqref{fullparam1} are considered instead, 
(ii) approximating the distribution of subgrid flux might present less complexity compared with the distribution of the net flux, 
(iii) decomposition \eqref{fbeslow} with \eqref{sgf} enforces mass conservation (at least on average) that is also achieved in application of flux difference scheme for $\mathbf{U}$ in the full system, 
(iv) same flux difference scheme is used for the effective equation as in the full model with parametrized subgrid fluxes considered as corrections;  this allows for better understanding of relative strengths
of flux terms and potential introduction of empirical corrections.

Replacing the sudgrid fluxes $G_{I+1/2}$ with a suitable parametrization $\tilde G_{I+1/2}$, 
the effective (or reduced) equation for large-scale variables becomes
\begin{align}
    \dfrac{d}{dt}U_I &= -\dfrac{F_{I + 1/2}-F_{I-1/2}}{n\Delta x} - \dfrac{\tilde G_{I+1/2} - \tilde G_{I-1/2}}{n \Delta x} + \rho^{U_I}. \label{eff}
\end{align}
Here, we assume that $\tilde G_{I+1/2}$ for each $I$ is a random variable, possibly conditioned on some neighboring resolved variables. Parametrization $\tilde G_{I+1/2}$ represent mechanism for effectively
sampling the distribution of $G_{I+1/2}$ and, thus, allowing for an efficient numerical 
simulation of the effective equation in \eqref{eff}.
In general, there are no restrictions on the form of $\tilde G_{I+1/2}$. 
In this paper we utilize a neural network to represent $\tilde G_{I+1/2}$.
In particular, we use generative adversarial network (GAN) to generate samples from the
distribution of $G_{I+1/2}$. Since the parametrization is inherently stochastic, we focus our attention of
reproducing the statistical properties of the resolved variables. Emphasis on statistical properties 
is, for instance, motivated by necessity to develop effective model for ensemble forecasts in 
climate and weather predictions.

\subsection{Forced Burgers' equation}
\label{sec:burgers}

Viscous Burgers equation is one of the well-studied nonlinear partial differential equations and
it is often used as a prototype model for turbulence \cite{burgers}.
We consider a stochastically forced Burgers equation over a periodic domain $[0,L]$ given by 
\begin{align}
    \dfrac{\partial u}{\partial t} +\dfrac{\partial}{\partial x}\left( \dfrac{u^2}{2} - \nu \dfrac{\partial u}{\partial x}\right) = \rho(x,t), \label{fbe}
\end{align}
where $\nu$ is the viscosity coefficient and $\rho$ is a large-scale stochastic forcing. To produce a system admitting the form given in (\ref{fullu}) we adopt the flux discretization used in \cite{kz}
and the discrete approximation for the flux becomes
\begin{align}
    F_{i+1/2} = \dfrac{1}{6}\left( u_{i+1}^2 + u_{i+1}u_i + u_i^2\right) - \dfrac{\nu}{\Delta x}\left(u_{i+1} - u_i\right). \label{fluxfbe}
\end{align}
Large-scale forcing is applied to low wavenumbers $k \in K$. The forcing term is chosen such that 
$\rho^{y_i} = 0$ for all $i$, and 
\begin{align*}
    \rho^{U_I} = \dfrac{A}{\sqrt{\Delta t}} \displaystyle\sum_{k \in K} \left[ \alpha_k \cos \left( 2\pi k\dfrac{n\Delta x}{L}\right) + \beta_k \sin \left( 2\pi k\dfrac{n\Delta x}{L}\right)\right], 
\end{align*}
where $\Delta t$ is the integration time step of the full model, $A$ is the amplitude, 
$\alpha_k$ and $\beta_k$ are independent random coefficients sampled from $N(0,1)$. 

For the Burgers' equation in \eqref{fbe} with the flux discretization \eqref{fluxfbe}
the self-interactions of the resolved variables and 
the subgrid flux in \eqref{sgf} become
\begin{align}
    F_{n(I+1)-1/2} := F_{I+1/2}\left( U_{I+1}, U_{I}\right) +  G_{I+1/2}\left( U_{I+1}, U_{I}, y_{I}^r, y_{I+1}^l\right),  \label{sgffb} 
\end{align}
where
\[
F_{I+1/2} = \dfrac{1}{6}\left( U_{I+1}^2 + U_{I+1}U_I + U_I^2\right) - \dfrac{\nu}{\Delta x}\left(U_{I+1} - U_I\right)
\]
and the subgrid flux depends on resolved variables $U_I$ and $U_{I+1}$, as well as on the unresolved modes
$y_{I}^r$ and $y_{I+1}^l$. Here $y_{I}^r$ and $y_{I+1}^l$ represent residuals located the right boundary of cell $I$ and left boundary of cell $I+1$, respectively.
Thus, for simpler notations, instead of using fine mesh index, we use $l$ and $r$, signifying  $left$ and $right$, to refer to location of the residuals with respect to its coarse cell index subscript. One essential thing to note about applying the averaging operator used in (\ref{aveop}) is that the resulting flux difference scheme admits a viscosity coefficient $n$ times larger than that in the full model. This is the direct consequence of applying the discrete averaging operator \eqref{aveop} to the full dynamics \eqref{fullu} with flux definition in \eqref{fluxfbe}.


The subgrid flux term can further be divided into subgrid contributions to the nonlinear part, 
$u^2/2$, and to the viscosity term, $\nu \partial_x u$. 
That is, we can represent the subgrid flux as 
\begin{align}
    G_{I+1/2}\left( U_{I+1}, U_{I}, y_I^r, y_{I+1}^l\right) = G_{I+1/2}^{(1)}\left( U_{I+1}, U_{I}, y_I^r, y_{I+1}^l\right) - \dfrac{\nu}{\Delta x}G_{I+1/2}^{(2)}\left( y_I^r, y_{I+1}^l\right).
    \label{Gsplit}
\end{align}
There are several reasons for such decomposition. First, 
terms on the right-hand side of \eqref{Gsplit} have different complexity - the first term involves quadratic interaction between resolved and unresolved modes, and the second term only involves linear relationship between unresolved modes. Second, we can easily investigate scalability of the stochastic parametrization with respect to varying $n$.

Given the discrete flux and forcing above, we use Runge-Kutta 3 time-stepping scheme 
for integrating the full model. We use the following parameters in our simulations 
\begin{center}
\begin{tabular}{ccccccc}
\hline $L$ & $N$ & $n$ & $\nu$ & $A$ & $K$  & $\Delta t$ \\
\hline 100 & 512 & 16 & 0.02 & $\sqrt{2} \times 10^{-2}$ & $\{1,2,3\}$ & 0.01 \\ \hline
\end{tabular}
\end{center}
We perform direct numerical simulations (DNS) of the full model in \eqref{fbe}
and generate long stationary time-series of $u_i(t)$. We then compute the resolved variables
using \eqref{aveop} and compute numerically their statistical properties, such as spectra, 
distribution, correlation function, and kurtosis. We then train a conditional WGAN to reproduce subgrid 
fluxes and perform simulations of the effective model in \eqref{eff}. We then compare statistical behavior
of resolved variables in the DNS and simulations of the effective model.

\subsection{WGAN parametrization}
\label{sec:wgan}
In this section, we present the structure of the network and describe the training procedure. 
we also describe data used for training and verification. 

\subsubsection{Network}


We use the conditional variant of WGAN as our network with the aim of generating samples dependent on the state of local averages. Moreover, our goal is to achieve a {\it local} parametrization, that is, produce approximate subgrid flux at each $I+1/2$ conditioned only on immediate neighboring local averages $U_I$ and $U_{I+1}$. Since the resolved and unresolved modes exhibit spatial homogeneity, it is sufficient to construct and train a generator using data for a particular cell $I$. In our case we consider $I=0$ and we use time series of $U_0$, $U_1$, $G_{1/2}^{(1)}$, and $G_{1/2}^{(2)}$ for training and testing. 

The generator accepts a vector $Z$ consisting of two random number  independently drawn from the uniform distribution on $[-1,1]$, and the conditions $U_0$ and $U_1$ at a specific time $t$. It outputs two values - $\tilde G_{1/2}^{(1)}$, and $\tilde G_{1/2}^{(2)}$. The discriminator takes values of $G_{1/2}^{(1)}$ and $G_{1/2}^{(2)}$, computed from reference set-up, as well as the generated $\tilde G_{1/2}^{(1)}$, and $ \tilde G_{1/2}^{(2)}$. It then decides how likely are the generated values drawn from the distribution of subgrid flux tendencies by giving "scores" to all the inputs. (Note that in vanilla GAN, discriminator gives probabilities of being "real" to inputs. Here we use "scores" because our training uses Wasserstein GAN. More details is presented in the next subsection.)  Both the generator and discriminator are fully-connected feed-forward networks with 3 hidden layers, 16 neurons in each layer and Leaky ReLU activation.

\begin{figure}[htbp]
\centering
\includegraphics[width=0.8\linewidth]{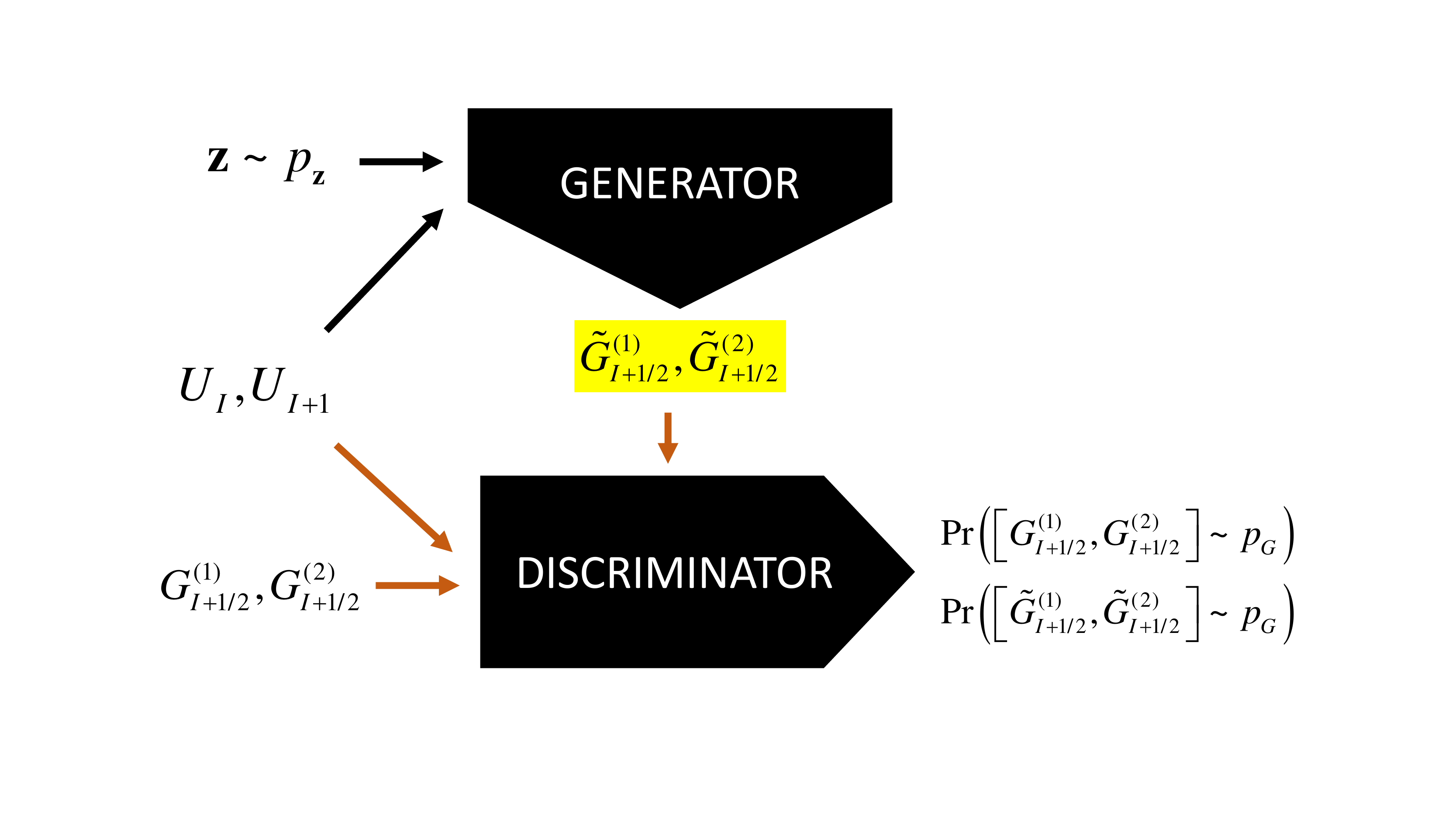}
\caption{Conditional GAN structure for subgrid flux parametrization}
\label{fig:pdf}
\end{figure}

\subsubsection{Training} 

We perform 
high-resolution numerical simulation of the full model and sample the solution every $50\Delta t$. After a burnout of 10,000 model time units, time-series of all four variables are 
($U_0$, $U_1$, $G_{1/2}^{(1)}$, and $G_{1/2}^{(2)}$)
are saved until $6 \times 10^5$ samples for each are obtained. The first $10^5$ is used for training and the rest is for validation. In training, we use Wassertein GAN with gradient penalty (WGAN-GP) since its loss function has more interpretability than that of the game-theoretic loss of vanilla GAN. The gradient penalty is included in the loss function to impose Lipschitz condition required for convergence of Wassertein distance between distributions. We refer to \cite{wgan, wgangp} for thorough discussion on this matter. The code for training and training data set is available at github. 

Stochastic gradient descent method is implemented for optimization with minibatch size of 400. Initial learning rate in Adam optimizer is set to $2 \times 10^{-5}$. During training, Wassertein distance between the distribution of "true" subgrid flux tendencies and the generated subgrid flux tendencies using the validation set is computed every epoch. Optimized weights and biases at epoch 100 are used to generate samples of subgrid flux tendencies in the reduced model \eqref{sgffb}.


\section{Results}
\label{sec:results}

Aside from bare truncation model (BTR) which integrates the reduced equation \eqref{eff} in time without any subgrid flux corrections, we also compare the performance of GAN parametrization with a polynomial regression with additive noise (POLY) similar to one used in \cite{wilks05}. This reduced model parametrizes subgrid scales by fitting a third-order polynomial with Gaussian additive noise estimated from regression residuals. The deterministic part of POLY parametrization is a multivariate polynomial for each $G^{(i)}_{1/2}$ ($i=1,2$) regressed against $U_0$ and $U_1$. Coefficient estimation was done in MATLAB. The same polynomial parametrization (with independent noise realizations) is applied for each coarse cell.  

\noindent 
\emph{Time integration:} For BTR, Runge-Kutta of order 3 (RK3) is used. For the other two reduced models (POLY and GAN), a splitting scheme is implemented. That is, the deterministic part of the subgrid flux tendency is integrated using RK3 while the stochastic term (additive noise or GAN-generated samples) is then added to the result in the manner similar to how Euler-Maruyama handles stochastic term in simulating SDEs. Time step used in simulations of reduced models
is equal to the time step in the numerical integration of the full model.

\subsection{Stationary statistics}
\label{sec5.1}

Statistical properties of trajectories are often used to access the 
performance of reduced models. To this end, we performed 
five long simulations with different realizations of forcing. All statistics are 
taken as ensemble and spatial averages (since the system is spatially homogeneous). 

\begin{figure}[h]
\centerline{
\includegraphics[scale=0.5]{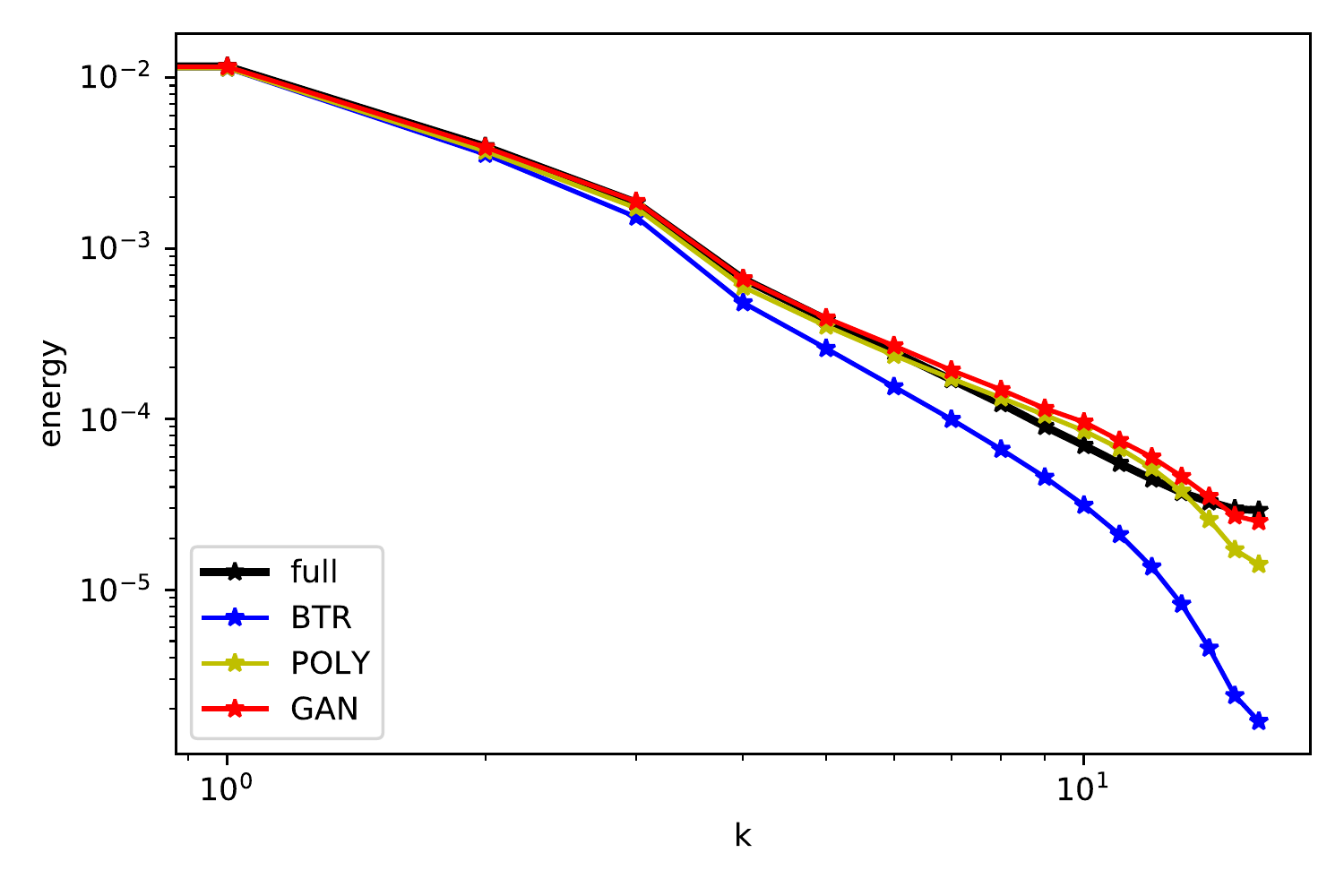}
\hspace{5pt}
\includegraphics[scale=0.5]{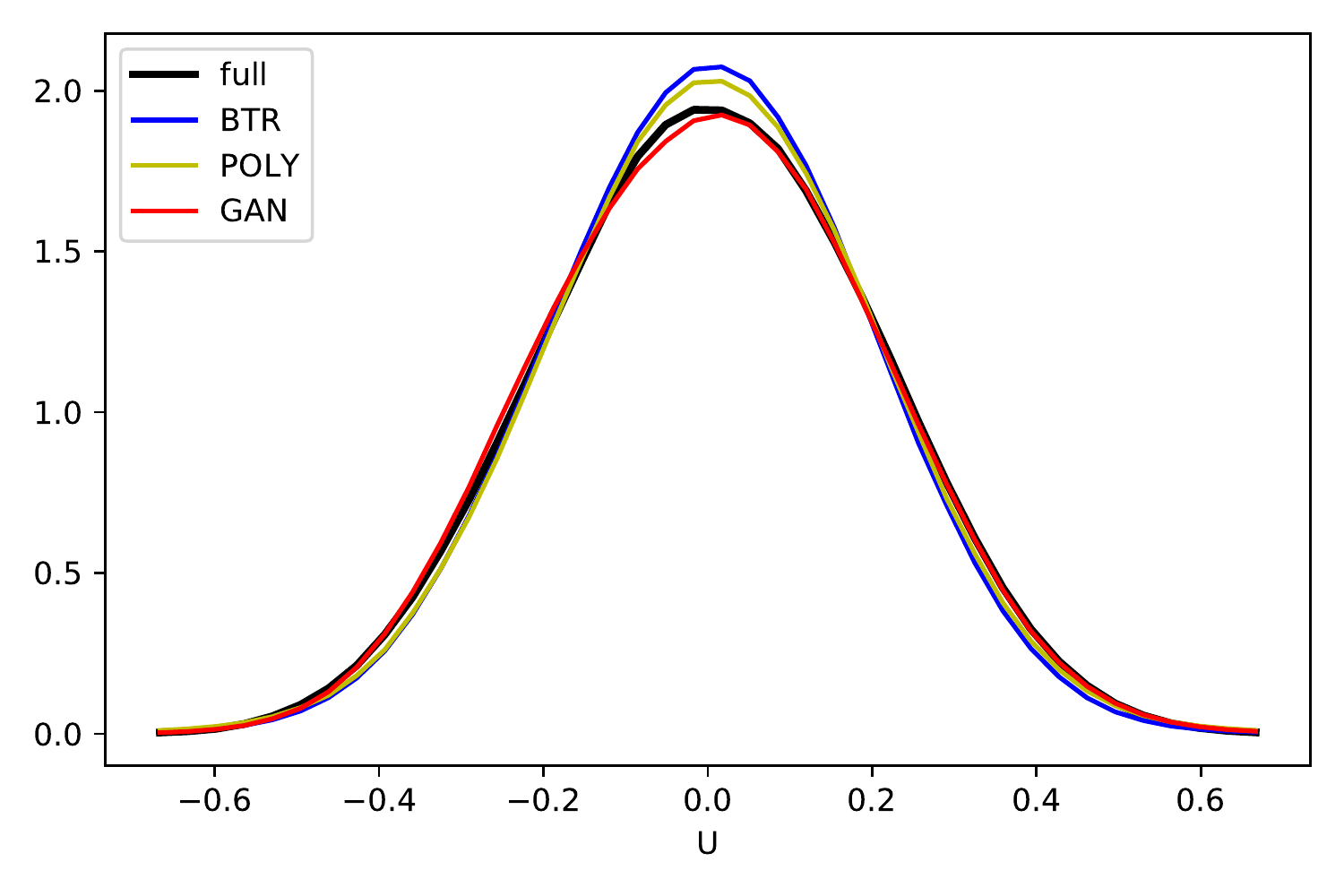}}
\caption{Energy spectrum (left) and pdfs (right) and of resolved variables $U_I$ in full 
model and reduced models.} \label{fig:pdfspec}
\end{figure}

Energy spectrum and probability density for the resolved variables in the full model and  different reduced models are presented in Figure \ref{fig:pdfspec}. 
Stationary variance and forth moment for $U_I$ are also presented in Table \ref{tab:1pt}.
Energy spectra clearly 
demonstrates that a sub-grid parametrization is necessary, since the bare-truncation (BTR)
underestimates the spectrum at high wavenumbers. The role of the parametrization is then 
to introduce corrections which mimic the sub-grid fluxes and reproduce the statistical 
properties more accurately. Both polynomial (POLY) and neural-network (GAN) 
reduced models reproduce energy spectrum quite accurately with GAN reduced model 
providing a slightly better approximation at the highest wavenumbers.         
Figure \ref{fig:pdfspec} shows that the pdf of the reduced variables is approximately normal.
It is evident that the shape is best resembled by the GAN parametrization. In addition, 
the GAN parametrization also reproduces one-point statistics (see Table \ref{tab:1pt})
slightly more accurately compared to the polynomial reduced model.
Mean values and third moments of reduced variables are omitted as all of them are very close to zero, similar to the DNS results. Because of higher viscosity and absence of any corrections, bare truncation is expected to underestimate the variance and hence results in a PDF which has higher values near the mean. Polynomial model provides some improvement for one-point statistics compared to the bare truncation, but still the pdf is slightly narrower than the results of the DNS.
\begin{table}[h]
    \centering
    \begin{tabular}{lccccc}
        \hline & Full &  BTR & POLY  & GAN \\ \hline
         Variance & 0.0393 & 0.0356 & 0.0381 & 0.0394 \\ \hline
         Fourth moment & 0.0043 & 0.0038 & 0.0046 & 0.0044 \\ \hline 
    \end{tabular}
    \smallskip
    \caption{Variance and fourth moments of resolved modes in full and reduced models.}
    \label{tab:1pt}
\end{table}

\begin{figure}[h]
\centerline{
\includegraphics[scale=0.5]{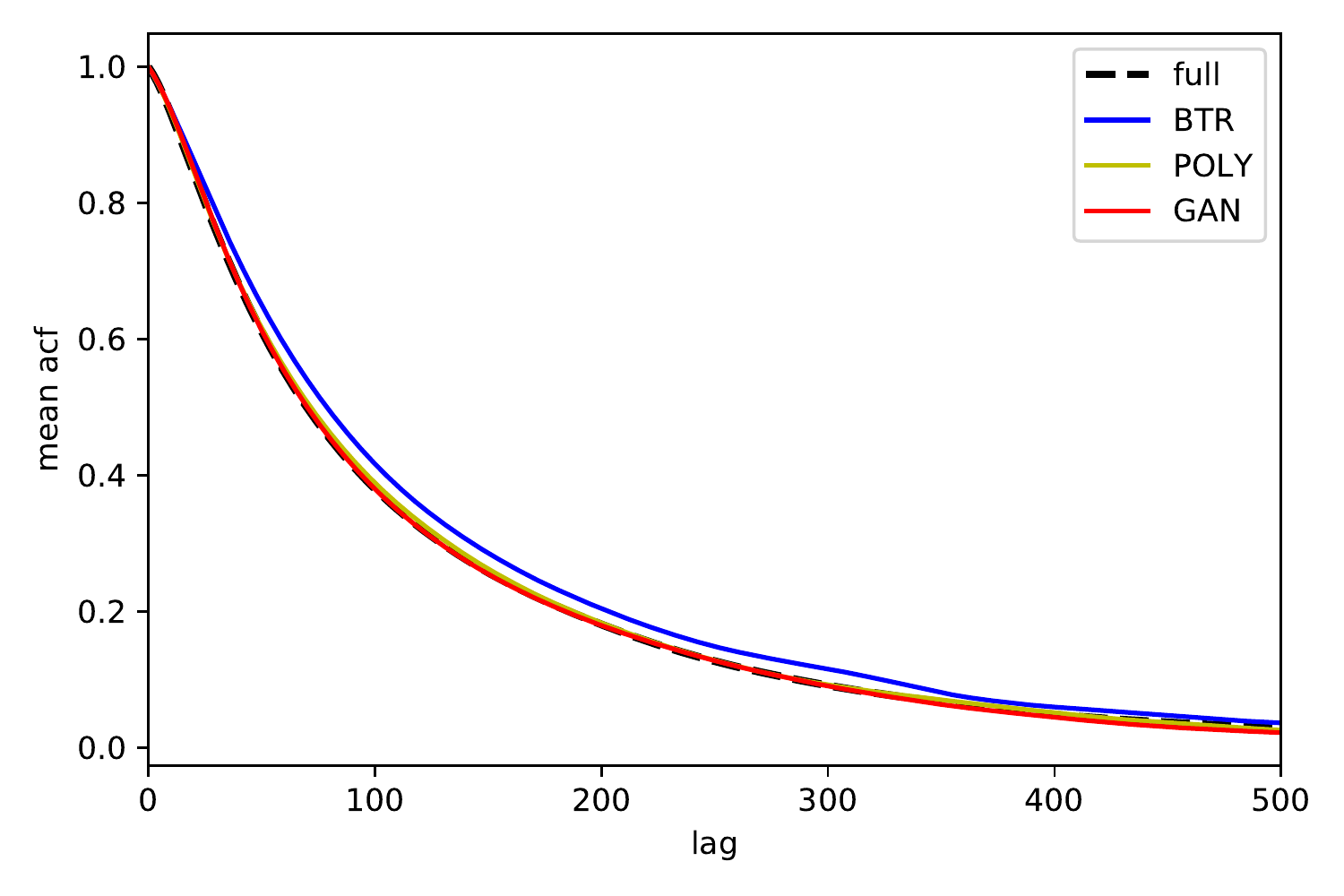}
\hspace{3pt}
\includegraphics[scale=0.55]{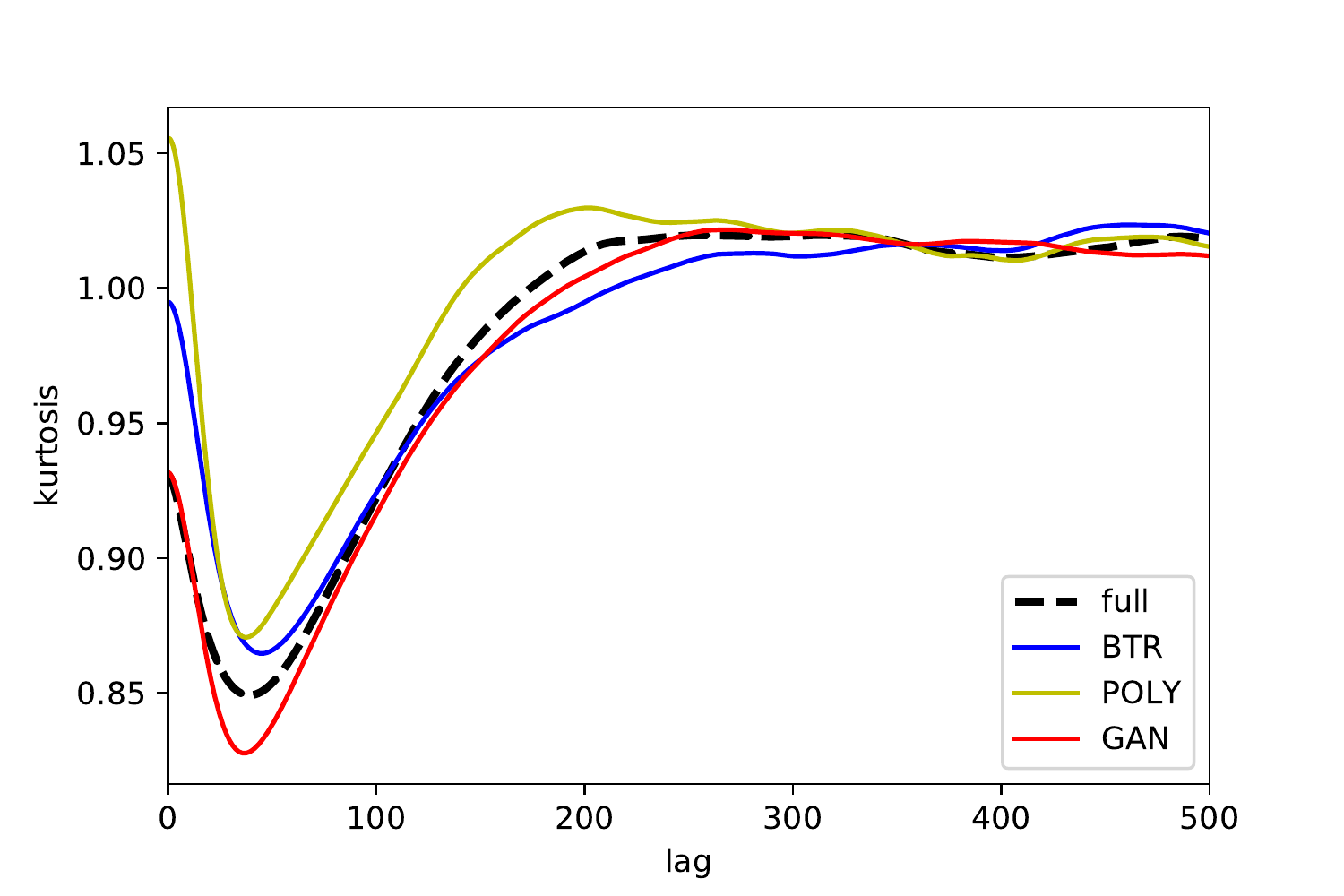}}
\caption{Two-point stationary statistics of reduced variables - correlation function (left) and kurtosis $K(s)$ (right) in full mode and reduced model.} \label{fig:ts}
\end{figure}

Typically, it is more challenging to reproduce two-point statistical properties, 
compared to the one-point stationary statistics. Therefore, we also compute 
two-time stationary statistics and compare performance of the three reduced models
with the results of DNS. Two-time stationary correlation function and kurtosis are depicted in Figure \ref{fig:ts}. Kurtosis is a measure of non-Gaussianity 
and it is given by
\[
K(s) = \dfrac{\langle U_i(t)^2 U_i(t+s)^2 \rangle}{\left(\langle U_i(t)^2 \rangle + 2\langle U_i(t)U_i(t+s) \rangle \right)^2}.
\]
The value of $K(s)$ is exactly one for Gaussian models.
All three reduced models reproduce stationary auto-correlation very well.
The GAN parametrization predicts the non-Gaussian features better compared to the polynomial and bare-truncation reduced model. Relative errors for $K(s)$ for the POLY model
are approximately 8-10\% compared to approximately 3\% relative errors for $K(s)$ in the 
GAN model. This is particularly evident for short lags.
Clearly, GAN reduced models performs very well in terms of capturing long-term statistical properties of local averages. Polynomial model also performs reasonably well. However, recall
that polynomial model includes additive Gaussian noise. Stationary statistics indicates
that local averages are almost Gaussian - we can see that the pdf is very close to a Gaussian distribution and kurtosis does not deviate from significantly from one. We can expect that 
polynomial reduced model will perform worse for stronger non-Gaussian models. On the other hand, 
due to the general nature of the noise in the GAN reduced model, we expect that this model
should perform adequately for non-Gaussian processes as well. This will be investigated in
subsequent papers.
Reduced models significantly accelerate computations. In particular, 
reduced GAN model is approximately three times faster compared with the DNS.

\subsection{Sensitivity to forcing}
\label{sec5.2}

It is important to develop parametrizations which can be easily adapted to different parameter and forcing regimes without additional re-training. Parametrization's ability to perform adequately 
with respect to changes in parameters/forcing/model resolution is an important practical factor.
Practical significance of parametrization is significantly enhanced if
it can perform well in a relatively wide parameter regime without re-fitting or re-training.
Therefore, we test the performance of the GAN parametrization with respect to increase in 
the magnitude of the forcing. In particular, we increase the magnitude of the forcing, $A$, 
by 10\% and 20\% and compare the performance of the GAN reduced model with DNS.
We would like to emphasize that the GAN reduced model is \emph{not} re-trained, i.e. 
we use GAN reduced model trained using data generated with the original forcing amplitude.


Polynomial model is excluded from 
the comparison here because the polynomial reduced model produced cases when the system's 
energy explodes.
With increased forcing amplitude, there are occurrences of local averages higher 
in magnitude than those in the data generated at the original forcing.
This leads to the incorrect estimation of sub-grid fluxes by the polynomial regression
and the polynomial model often diverges.
The most likely reason for a poor performance of the polynomial reduced model is
inadequate estimation of sub-grid fluxes in regions with high gradients.
When an under- or overestimation of subgrid flux corrections occurs in polynomial model, near-discontinuties in regions with high gradients may not be sufficiently "tamed" by viscosity which leads to more instances of non-physical shocks or oscillations.  
Although this event can be hard to disprove in the case of GAN, it was observed that GAN did not encounter this problem. 

\begin{figure}[h]
\centering
\includegraphics[scale=0.5]{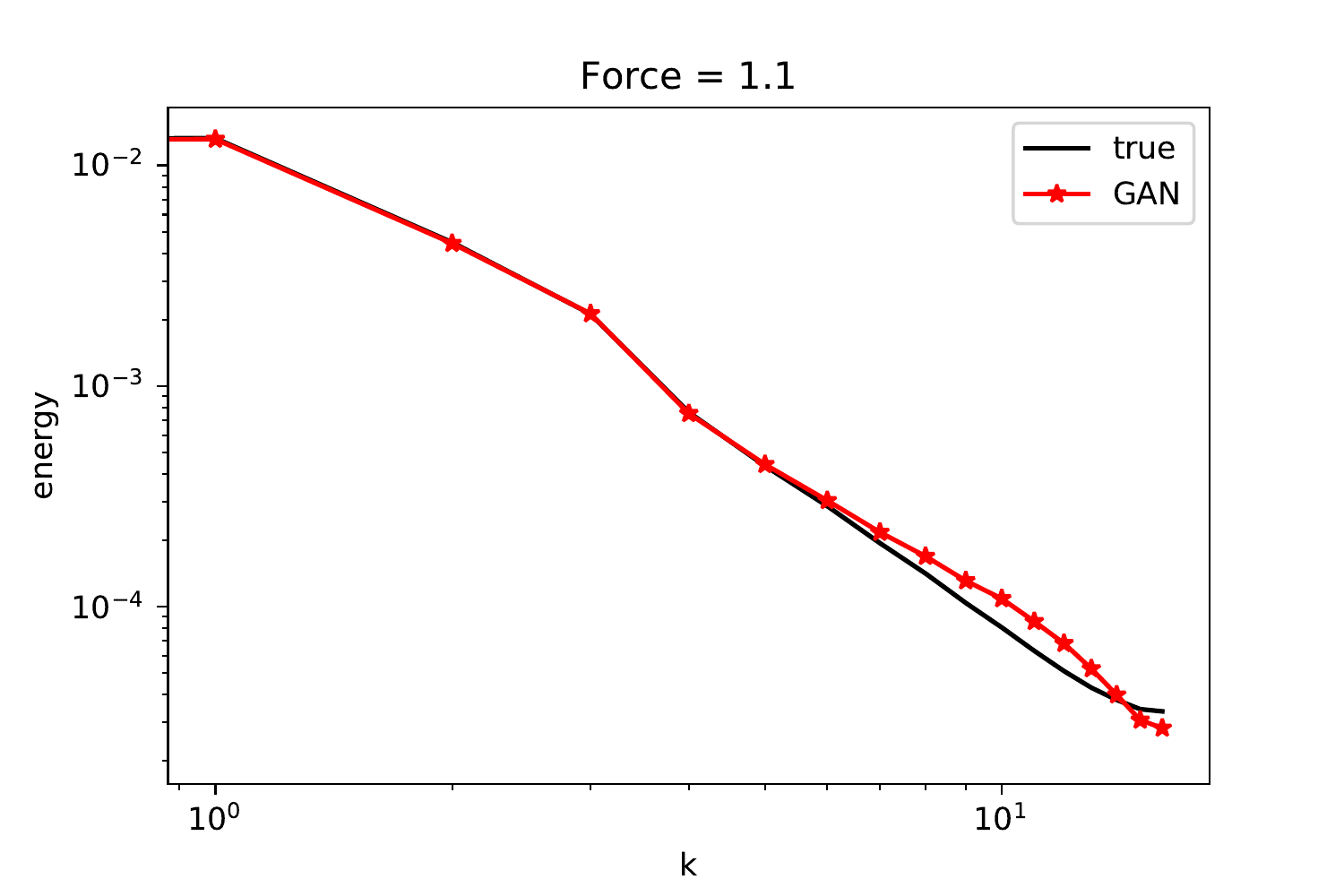}
\hspace{5pt}
\includegraphics[scale=0.5]{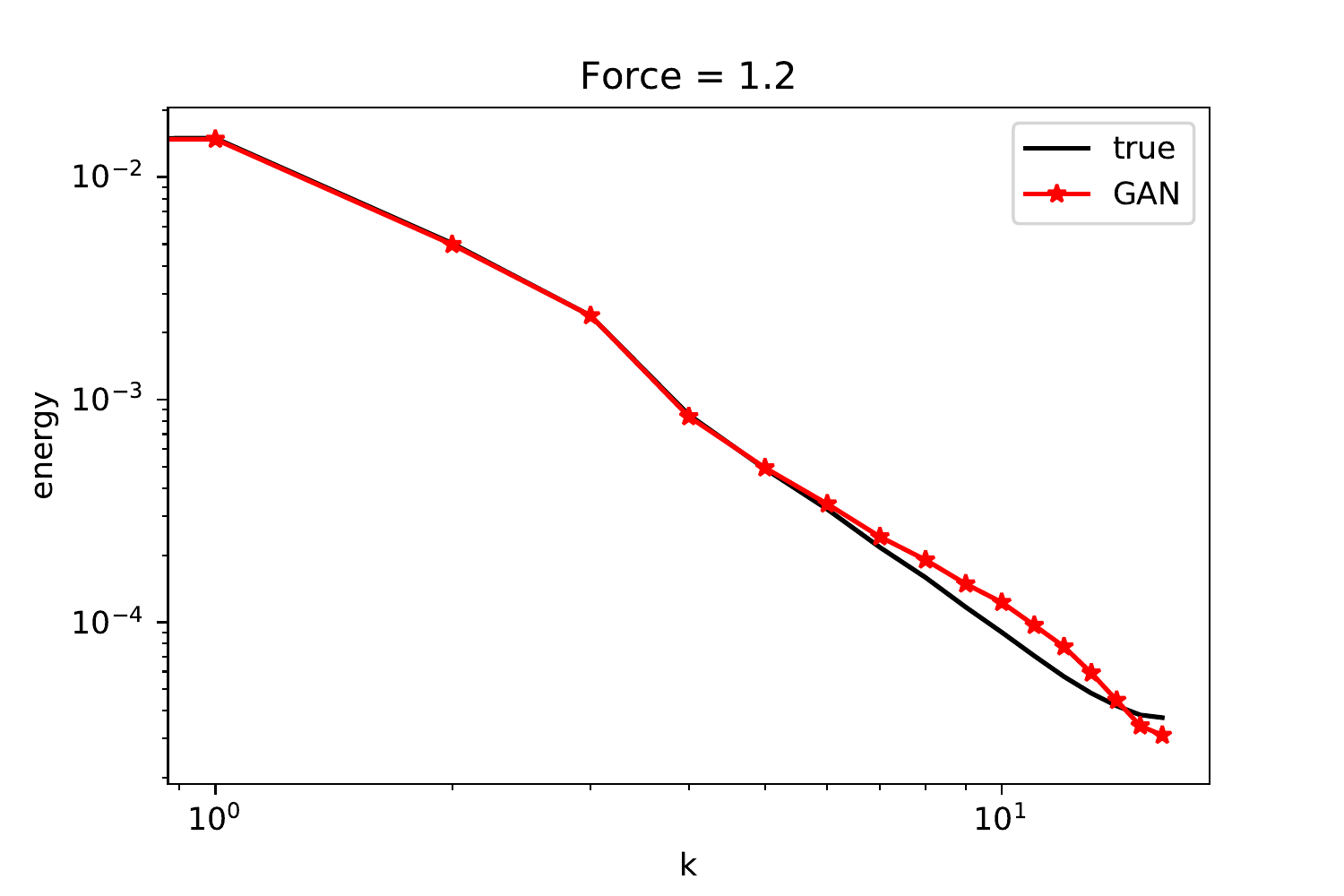}
\caption{Energy spectrum of resolved variables in simulations with forcing amplitude at 110\% (left) and 120\% (right).} \label{fig:specf}
\end{figure}

Numerical results for comparison of the GAN reduced model with DNS in the regime with
increased forcing are presented in Figure \ref{fig:specf} and Table \ref{tab:1ptf}.
The GAN reduced model performs very well in both cases.
We would like to point out that increase in forcing magnitude by 10\% and 20\%
results in the increase in variance of the resolved variables by approximately
12\% and 26\%, respectively. One-point statistical quantities are reproduced perfectly 
by the GAN reduced model. Two-point statistics is also reproduced very-well, 
including the kurtosis $K(s)$ and these results are not presented here only for the brevity 
of presentation.
\begin{table}[h]
    \centering
    \begin{tabular}{lcc}
        \hline & Variance & Fourth moment  \\ \hline
         DNS (10\%) & 0.0441 & 0.0055  \\ \hline
         GAN  (10\%) & 0.0440 & 0.0055  \\ \hline \hline
         DNS (20\%) & 0.0496 & 0.0069  \\ \hline
         GAN  (20\%) & 0.0495 & 0.0070  \\ \hline
    \end{tabular}
    \smallskip
    \caption{Variance and fourth moments of resolved modes in simulations with increased forcing by 10\% and 20\%.}
    \label{tab:1ptf}
\end{table}

Results in this section demonstrate that the
GAN reduced model is able to reproduce the behavior of resolved variables in DNS
with different forcing amplitude. We would like to point out that changes in forcing 
amplitude resulted in substantial increase in the variability of resolved variables. 
Therefore, the GAN reduced model proves capable of extrapolating subgrid flux tendencies 
outside of the parameter regime where it was trained.

\subsection{Scalability}
\label{sec5.3}

Another important practical property of parametrizations is scalability.
Scalability reflects the ability to utilize the same parametrization without re-training in a reduced model with different spatial (and possibly temporal) resolution. 
Typically, if the spatial resolution is changed 
(e.g. coarser or finer mesh than originally designed is used), 
we would like to use the same parametrization scheme (possibly properly scaled) to 
perform simulations of reduced models on a new mesh without re-training or re-fitting.
Resolution decrease is motivated by necessity to run fast crude simulations 
in, for example, ensemble predictions, and increase in resolution is motivated by
technological advances and future increase in computational capabilities. Thus, in the 
future it should be possible to run reduced models with higher spatial resolution, and it is important to verify that stochastic parametrizations utilized in reduced models can adequately respond without re-training. 

Here we demonstrate that simple linear scaling in GAN parametrization is sufficient 
to adapt GAN reduced model with respect to changes in mesh size and
perform simulations with a different spatial resolution. Of course, we cannot expect that 
the same parametrization would perform adequately if the spatial resolution is changed drastically. 
Therefore, we test several cases  - when the mesh becomes twice finer and
two and four times coarser.
In particular, we consider the following setup. Recall, that we denote 
the averaging window as $n$ (see \eqref{aveop}).
If the averaging window is changed to $n'$, then we use a linear scaling and 
the generated subgrid fluxes (output of GAN)
is multiplied by a factor $n/n'$. This linear scaling is motivated by the 
application of a systematic stochastic model-reduction 
strategy to finite-difference discretization of the Burgers-Hopf equation 
\cite{dta12,dat12} and the shallow-water model \cite{zdat18}. 

The reference setup in section \ref{sec5.1} corresponds to $n=16$ and here we
consider cases $n' = 4$, $n'=8$, and $n'=32$.
Second and fourth moments in simulations with $n=16$ and $n'=4, 8, 32$ are
presented in Table \ref{tab:1pts}. In addition, in this Table we also present results for 
simulations with non-scaled (NS) GAN parametrizations. 
We can see that the scaling introduced earlier improves the performance of the reduced model 
for all cases $n'=4$, $8$, $32$.
Numerical results for $n'=4$ and $n'=8$ are very close to the 
results for $n=16$ and, thus, scaling of the GAN parametrization does not produce a large effect.
On the other hand, numerical results for $n'=32$ clearly indicate that 
scaling of the GAN parametrization improves the performance of the reduced model.
This is particularly important for simulations with the coarser resolution 
$n'=32$. 
We would like to point out that variance 
of reduced variables does not increase significantly for cases $n' = 4$ and $n'=8$ 
(only 2.7\% and 1.5\% relative increase, respectively), but for $n'=32$ variance is reduced by approximately 8\% compared with $n=16$. This is likely the reason why the reduced model 
for $n'=32$ is more sensitive to the scaling of GAN parametrization compared to simulations with
$n'=4$ and $n'=8$.
Two point stationary statistics are also reproduced very well by reduced models 
with scaled GAN parametrization.
Overall, there is a good agreement between the direct numerical simulations and 
simulations of the  reduced model with scaled GAN parametrization. 
In addition to results presented in 
Table \ref{tab:1pts}, utilizing scaled GAN parametrization also improves the
shape of the spectrum for high wavenumbers for $n'=4$ and $n'=8$ (see \cite{Jericthesis}).
Simulations with unscaled GAN parametrization with $n'=32$ completely fail to 
reproduce stationary spectrum and probability distribution of resolved variables \cite{Jericthesis}. 
These results are not presented here only for the brevity of 
presentation.
\begin{table}[h]
    \centering
    \begin{tabular}{lcc|cc}
        \hline & Variance & Rel. error & Fourth momen & Rel. error   \\ \hline
         DNS ($n'=4$) & 0.04044 &  & 0.004625 &  \\ 
         GAN-NS  ($n'=4$) & 0.03935 & (2.7\%) & 0.004455 & (3.7\%)\\
         GAN-S  ($n'=4$) & 0.03992 & (1.3\%) & 0.004491 & (2.9\%)\\  \hline
         DNS ($n'=8$) & 0.03993 &  & 0.004509 &  \\ 
         GAN-NS  ($n'=8$) & 0.03842 & (3.8\%) & 0.004253 & (5.7\%)\\
         GAN-S  ($n'=8$) & 0.03942 & (1.3\%) & 0.004383 & (2.8\%)\\  \hline
         DNS ($n=16$) & 0.03934 &  & 0.004347 & \\ 
         GAN  ($n=16$) & 0.03946 & 0.3\% & 0.004379 & 0.7\%\\  \hline
         DNS ($n'=32$) & 0.03616 & & 0.003689 &  \\
         GAN-NS  ($n'=32$) & 0.05927 & 63.9\% & 0.007950 & 115.5\% \\
         GAN-S  ($n'=32$) & 0.03717 & 2.8\% & 0.004006 & 8.6\% \\ \hline
    \end{tabular}
    \smallskip
    \caption{Variance and fourth moments of resolved modes in Direct Numerical Simulations and 
    simulations of the reduced model with different resolutions and GAN parametrization. GAN-S and GAN-NS denote scaled and non-scaled version of the GAN parametrization trained with $n=16$, respectively.}
    \label{tab:1pts}
\end{table}

\section{Conclusions}
\label{sec:conc}
A stochastic subgrid flux parametrization is developed here using conditional GAN. We demonstrated
that Generative Adversarial Networks can be successfully trained using the Wasserstein metric to reproduce
subgrid corrections for fluxes of resolved variables. The WGAN version of the Adversarial Networks
significantly improves conversion during the training process.
In addition, to estimate the subgrid flux between two neighboring cells, 
the GAN is conditioned on the current value of the solution in these two cells. 
This conditioning is essential for reproducing the subgrid fluxes.
Resolved variables are defined here as local spatial averages.
This makes the reduced model local in space, i.e. only two neighboring cells are coupled by a 
GAN parametrization which estimates the flux between these two cells. Thus, if the model is 
non-homogenious in space, the GAN parametrization can potentially be trained separately for 
different spatial regions.

Our aim here is to reproduce statistical properties of long-term turbulent simulations.
In particular, we compare spectra, moments, and two-point temporal statistical quantities.
We would like to point out that we simulate the reduced model for a long time and the reduced model
does not blow up; all runs of the reduced model produce stable trajectories.
The reduced model with GAN parametrization reproduces statistical properties of the full
model quite well. We illustrate our approach using the forced Burgers-Hopf model.
In addition, we investigate numerically the performance of the GAN parametrization with respect to
changes in forcing and different spatial resolutions. To this end, we utilize the original 
parametrization to simulate reduced models with 10\% and 20\% increase in forcing.
We also introduce linear scaling for the GAN parametrization to compute reduced models with
different spatial resolutions. We demonstrate that this linear scaling is essential for 
the reduced model to accurately reproduce statistical properties of solutions.
We would like to emphasize that the GAN parametrization is not re-trained on the new data
(e.g. simulations with increased forcing or changed resolution). Therefore, we demonstrate that
the reduced models with the GAN paramterization can be used in a wide range of regimes without
re-training the neural network.

The approach developed here is particularly suitable for 
finite-volume (or finite-difference) discretizations of partial differential equations since 
resolved variables are defined as local spatial averages and subgrid fluxes are represented using
corresponding averages of the right-hand side of the equation. We demonstrated applicability of this
approach for a prototype nonlinear model which exhibits complex turbulent behavior. Thus, this approach stands ready to be applied in more complex models of fluid dynamics. This will be investigated in subsequent papers.

\section{Acknowledgments}

This research was partially supported by the grants ONR N00014-17-1-2845 and NSF-DMS 1620278.


\begin{thebibliography}{10}

\bibitem{Jericthesis}
J.~S. Alcala.
\newblock {\em Subgrid-scale parametrization of unresolved processes}.
\newblock PhD thesis, University of Houston, 2021.

\bibitem{wgan}
M.~Arjovsky, S.~Chintala, and L.~Bottou.
\newblock Wassertein gan.
\newblock {\em arXiv:1701.07875v3}, 2017.

\bibitem{bb}
N.~D. Brenowitz and C.~S. Bretherton.
\newblock Spatially extended tests of a neural network parametrization trained
  by coarse-graining.
\newblock {\em Journal of Advances in Modeling Earth Systems.}, 11:2728–2744,
  2019.

\bibitem{burgers}
J.M. Burgers.
\newblock A mathematical model illustrating the theory of turbulence.
\newblock {\em Adv. Appl. Mech.}, 1:171--179, 1948.

\bibitem{nn2}
A.~Chattopadhyay, E.~Nabizadeh, and P.~Hassanzadeh.
\newblock Analog forecasting of extreme-causing weather patterns using deep
  learning.
\newblock {\em arXiv:1907.11617}, 2019.

\bibitem{chh09}
A.~J. Chorin and O.~H. Hald.
\newblock {\em Stochastic Tools in Mathematics and Science}.
\newblock Springer, 2009.

\bibitem{Chorin2968}
A.~J. Chorin, O.~H. Hald, and R.~Kupferman.
\newblock Optimal prediction and the mori{\textendash}zwanzig representation of
  irreversible processes.
\newblock {\em Proceedings of the National Academy of Sciences},
  97(7):2968--2973, 2000.

\bibitem{Chorin2002239}
A.~J. Chorin, O.~H. Hald, and R.~Kupferman.
\newblock Optimal prediction with memory.
\newblock {\em Physica D: Nonlinear Phenomena}, 166(3):239 -- 257, 2002.

\bibitem{chorin15}
A.~J. Chorin and F.~Lu.
\newblock Effects of stochastic parametrization in the lorenz '96 system.
\newblock {\em Proc. Natl. Acad. Sci. U.S.A.}, 112(32):9804--9809, 2015.

\bibitem{markov1}
D.~T. Crommelin and E.~Vanden-Eijnden.
\newblock Subgrid-scale parametrization with conditional markov chains.
\newblock {\em J. Atmos. Sci.}, 65:2661--2675, 2008.

\bibitem{dat12}
S.~I. Dolaptchiev, U.~Achatz, and I.~Timofeyev.
\newblock Stochastic closure for local averages in the finite-difference
  discretization of the forced {Burgers} equation.
\newblock {\em Theor. Comput. Fluid Dyn.}, 27(3-4):297--317, 2013.

\bibitem{dta12}
S.~I. Dolaptchiev, I.~Timofeyev, and U.~Achatz.
\newblock Subgrid-scale closure for the inviscid {Burgers-Hopf} equation.
\newblock {\em Comm. Math. Sci.}, 11(3):757--777, 2013.

\bibitem{bams}
J.~Berner et~al.
\newblock Stochastic parametrization: Towards a new view of weather and climate
  models.
\newblock {\em Bull. Amer. Meteor. Soc.}, 98:565--588, 2017.

\bibitem{sppt2}
T.~Palmer et~al.
\newblock Stochastic parametrization and model uncertainty. technical report
  no. 596.
\newblock {\em ECMWF, Reading UK}, 2009.

\bibitem{low2}
C.~Franzke and A.~Majda.
\newblock Low-order stochastic mode reduction for a prototype atmospheric gcm.
\newblock {\em J. Atmos. Sci}, 63:457--479, 2006.

\bibitem{low1}
C.~Franzke, A.~Majda, and E.~Vanden-Eijnden.
\newblock Low-order stochastic mode reduction for a realistic barotropic model
  climate.
\newblock {\em J. Atmos. Sci}, 62:1722--1745, 2005.

\bibitem{frederiksen97}
J.~S. {Frederiksen} and A.~G. {Davies}.
\newblock {Eddy Viscosity and Stochastic Backscatter Parameterizations on the
  Sphere for Atmospheric Circulation Models.}
\newblock {\em Journal of Atmospheric Sciences}, 54:2475--2492, October 1997.

\bibitem{Frederiksen2013}
J.~S. Frederiksen, T.~J. O'Kane, and M.~J. Zidikheri.
\newblock {Subgrid modelling for geophysical flows}.
\newblock {\em Philosophical Transactions of the Royal Society A: Mathematical,
  Physical and Engineering Sciences}, 371(1982):20120166--20120166, 2012.

\bibitem{gan}
Ian~J. Goodfellow, Jean Pouget-Abadie, Mehdi Mirza, Bing Xu, David
  Warde-Farley, Sherjil Ozair, Aaron Courville, and Yoshua Bengio.
\newblock Generative adversarial networks.
\newblock {\em arXiv:1406.2661}, 2014.

\bibitem{wgangp}
I.~Gulrajani, F.~Ahmed, M.~Arjovsky, V.~Dumoulin, and A.~Courville.
\newblock Improved training of wassertein gans.
\newblock {\em arXiv:1704.00028v3}, 2017.

\bibitem{hassel}
K.~Hasselman.
\newblock Stochastic climate models. part I. theory.
\newblock {\em Tellus}, 28:473--485, 1976.

\bibitem{hevd10}
C.~Hijon, P.~Espanol, E.~Vanden-Eijnden, and R.~Delgado-Buscalioni.
\newblock Mori-Zwanzig formalism as a practical computational tool.
\newblock {\em Faraday Discuss.}, 144:301–302, 2010.

\bibitem{ganlorenz}
D.~J.~Gagne II, H.~M. Christensen, A.~C. Subramanian, and A.~H. Monahan.
\newblock Machine learning for stochastic parameterization: Generative
  adversarial networks in the {Lorenz '96} model.
\newblock {\em arXiv:1909.04711}, 2019.

\bibitem{markov2}
K.~Nimsaila K. and I.~Timofeyev.
\newblock Markov chain stochastic parametrizations of essential variables.
\newblock {\em SIAM Mult. Mod. Simul.}, 8(5):2079--2096, 2010.

\bibitem{kkg05}
S.~Kravtsov, D.~Kondrashov, and M.~Ghil.
\newblock Multilevel regression modeling of nonlinear processes: derivation and
  applications to climatic variability.
\newblock {\em J. Climate}, 18:4404--4424, 2005.

\bibitem{markov3}
F.~Kwasniok.
\newblock Data-based stochastic subgrid-scale parametrisation: an approach
  using cluster weighted modeling.
\newblock {\em Phil. Trans. R. Soc. A}, 370:1061--1086, 2012.

\bibitem{narmax2}
F.~Lu, K.~K. Lin, and A.~J. Chorin.
\newblock Data-based stochastic model reduction for the
  {Kuramoto–Sivashinsky} equation.
\newblock {\em Physica D}, 340:46--57, 2017.

\bibitem{mtv01}
A.~Majda, I.~Timofeyev, and E.~Vanden-Eijnden.
\newblock A mathematical framework for stochastic climate models.
\newblock {\em Commun. Pure Appl. Math}, 54:891--974, 2001.

\bibitem{mtv03}
A.~Majda, I.~Timofeyev, and E.~Vanden-Eijnden.
\newblock Systematic strategies for stochastic mode reduction in climate.
\newblock {\em J. Atmos. Sci.}, 60:1705--1722, 2003.

\bibitem{seam1}
A.~Majda, I.~Timofeyev, and E.~Vanden-Eijnden.
\newblock Stochastic models for selected slow variables in large deterministic
  systems.
\newblock {\em Nonlinearity}, 19:769--794, 2006.

\bibitem{gordw}
P.~A. O'Gorman and J.~G. Dwyer.
\newblock Using machine learning to parameterizemoist convection: Potential for
  modeling of climate, climate change, and extreme events.
\newblock {\em Journal of Advances in Modeling Earth Systems.}, 10:2548--2563,
  2018.

\bibitem{rasp}
S.~Rasp, M.~S. Pritchard, and P.~Gentine.
\newblock Deep learning to represent subgrid processes in climate models.
\newblock {\em PNAS}, 115(59):9684--9689, 2018.

\bibitem{resseguier20}
V.~Resseguier, L.~Li, G.~Jouan, P.~D\'erian, E.~M\'emin, and C.~Bertrand.
\newblock New trends in ensemble forecast strategy: uncertainty quantification
  for coarse-grid computational fluid dynamics.
\newblock {\em Archives of Computational Methods in Engineering}, In
  press:1--82 (hal--02558016), 2020.

\bibitem{sppt1}
G.~Shutts and T.~Palmer.
\newblock Convective forcing fluctuations in a cloud-resolving model: relevance
  to the stochastic parameterization problem.
\newblock {\em J. Clim.}, 20:187--202, 2007.

\bibitem{maxent1}
W.~T.~M. Verkley.
\newblock A maximum entropy approach to the problem of parametrization.
\newblock {\em Q. J. R. Meteorol. Soc.}, 137:1872--1886, 2011.

\bibitem{Verkley2016}
W.~T.~M. Verkley, P.~C. Kalverla, and C.~A. Severijns.
\newblock {A maximum entropy approach to the parametrization of subgrid
  processes in two-dimensional flow}.
\newblock {\em Quarterly Journal of the Royal Meteorological Society},
  142(699):2273--2283, 2016.

\bibitem{maxent2}
W.~T.~M. Verkley, P.~C. Kalverla, and C.~A. Severijns.
\newblock Amaximum entropy approach to the parametrization of subgrid processes
  in two-dimensional flow.
\newblock {\em Q. J. R. Meteorol. Soc.}, 142:2273--2283, 2016.

\bibitem{nn1}
P.~R. Vlachas, J.~Pathak, B.~R. Hunt, T.~P. Sapsis, M.~Girvan, E.~Ott, and
  P.~Koumoutsakos.
\newblock Forecasting of spatio-temporal chaotic dynamics with recurrent neural
  networks: a comparative study of reservoir computing and backpropagation
  algorithms.
\newblock {\em arXiv:1910.05266}, 2019.

\bibitem{wilks05}
D.S. Wilks.
\newblock Effects of stochastic parametrization in the {Lorenz} '96 system.
\newblock {\em Quart. J. Roy. Meteor. Soc.}, 131:389--407, 2005.

\bibitem{resp1}
J.~Wouters, S.~I. Dolaptchiev, V.~Lucarini, and U.~Achatz.
\newblock Parameterization of stochastic multiscale triads.
\newblock {\em Nonlinear Processes in Geophysics.}, 23:435--445, 2016.

\bibitem{kz}
N.~J. Zabusky and M.~D. Kruskal.
\newblock Interaction of “solitons” in a collisionless plasma and the
  recurrence of initial states.
\newblock {\em Phys. Rev. Lett.}, 15:240--243, 1965.

\bibitem{zdat18}
M.~Zacharuk, S.~I. Dolaptchiev, U.~Achatz, and I.~Timofeyev.
\newblock Stochastic subgrid-scale parameterization for one-dimensional shallow
  water dynamics using stochastic mode reduction.
\newblock {\em Q.J.R. Meteorol. Soc.}, 144(715):1975--1990, 2018.

\end{thebibliography}

\end{document}